\documentclass[aoas]{imsart}

\RequirePackage{amsthm,amsmath,amsfonts,amssymb}
\RequirePackage[authoryear]{natbib}
\usepackage{hyperref}
\usepackage[utf8]{inputenc}
\usepackage{graphicx}
\usepackage{color}
\usepackage{lineno}
\usepackage{multirow}
\usepackage{rotating}
\usepackage{arydshln}

\startlocaldefs
\endlocaldefs

\begin{document}

\begin{frontmatter}
\title{Conditional autoregressive models fused with random forests to improve small-area spatial prediction}
\runtitle{The CAR-Forest spatial prediction algorithm}

\begin{aug}
\author[A]{\fnms{Cara}~\snm{MacBride}\ead[label=e1]{2392494m@student.gla.ac.uk}},
\author[A]{\fnms{Vinny}~\snm{Davies}\ead[label=e2]{vinny.davies@glasgow.ac.uk}\orcid{0000-0003-1896-8936}}

\and
\author[A]{\fnms{Duncan}~\snm{Lee}\ead[label=e3]{Duncan.Lee@glasgow.ac.uk}\orcid{0000-0002-6175-6800}}
\address[A]{School of Mathematics and Statistics, University of Glasgow, Glasgow, G12 8SQ, Scotland.\printead[presep={,\ }]{e1,e2,e3}}
\end{aug}

\begin{abstract}
In areal unit data with missing or suppressed data, it desirable to create models that are able to predict observations that are not available. Traditional statistical methods achieve this through Bayesian hierarchical models that can capture the unexplained residual spatial autocorrelation through conditional autoregressive (CAR) priors, such that they can  make predictions at geographically related spatial locations. In contrast, typical machine learning approaches such as random forests ignore this residual autocorrelation, and instead base predictions on complex non-linear feature-target relationships. In this paper, we propose CAR-Forest, a novel spatial prediction algorithm that combines the best features of both approaches by fusing them together. By iteratively refitting a random forest combined with a Bayesian CAR model in one algorithm, CAR-Forest can incorporate flexible feature-target relationships while still accounting for the residual spatial autocorrelation. Our results, based on a Scottish housing price data set, show that CAR-Forest outperforms Bayesian CAR models, random forests, and the state-of-the-art hybrid approach, geographically weighted random forest, providing a state-of-the-art framework for small-area spatial prediction.
\end{abstract}


\end{frontmatter}

\section{Introduction}
Spatial areal unit data are prevalent in ecology (\citealp{brewer2007}), economics (\citealp{kawabata2022}), and epidemiology (\citealp{lee0223}), and the aims of modelling these data include hotspot identification (\citealp{knorrheld2000}), boundary detection (\citealp{lee2021}), ecological regression (\citealp{wang2022}), and the quantification of spatial inequalities (\citealp{jack2019}). Unlike for point-level data spatial prediction is not normally the inferential goal, because there is one data value for each areal unit and hence nothing to predict. However,  areal unit data sometimes contain missing values, making spatial prediction an important methodological challenge. These missing values could be caused by the observed value not existing, not being measured or being suppressed, the latter occurring because it may disclose the identity of individuals. Here, we model median property prices at a small-area scale in Scotland, and these data are only publicly released if 5 or more properties sold in a year, leading to around 9\% of the small areas having missing values. 

Area unit data are typically modelled in a Bayesian hierarchical setting, where the mean  function is represented by a linear combination of available features and a set of random effects. The latter capture any residual spatial autocorrelation in the data after feature adjustment, and are typically assigned a conditional autoregressive (CAR) prior distribution (\citealp{besag1991}). In contrast,  machine learning algorithms are the state of the art approach to a-spatial prediction, with examples including random forests (\citealp{breiman2001}) and gradient boosting machines (\citealp{friedman2001b}). These algorithms model the relationship between each feature and the target variable as a complex non-linear function, typically leading to improved predictive performance compared to simpler linear models. These competing paradigms thus utilise different aspects of spatial areal unit data to make predictions,  with machine learning algorithms utilising complex non-linear feature-target relationships and ignoring residual spatial autocorrelation, while Bayesian hierarchical CAR models capture this autocorrelation at the expense of simpler feature-target relationships. 

The use of machine learning in spatial statistics is a growing research area, with \cite{berrocal2020} and \cite{credit2022} comparing the predictive performance of traditional spatial statistical models and machine learning algorithms. A small number of hybrid methodologies have also been proposed that fuse these two approaches, including the random forest regression Kriging (RFRK, \citealp{hengl2015}) and random forest generalised least squares (RF-GLS, \citealp{saha2023}) algorithms for point-level data. For areal unit data \cite{xia2021} and \cite{soltani2022} incorporated spatially lagged features in tree-based machine learning models, while \cite{georganos2021} proposed a geographically weighted random forest (GRF) algorithm that fits a separate local random forest for each areal unit using only nearby data points. However, unlike the point-level RFRK and RF-GLS algorithms, these areal unit methods do not explicitly allow for spatial autocorrelation. 

Therefore, this paper proposes an iterative prediction algorithm for areal unit data called \texttt{CAR-Forest}, which is a novel fusion of conditional autoregressive models and random forests. The algorithm incorporates  flexible feature-target relationships via a random forest and residual spatial autocorrelation via a Bayesian CAR model, and iteratively re-fits each component based on the current value of the other. The total number of iterations is one of the tuning parameters of the algorithm, which are optimised via a 10-fold cross validation procedure. This methodology is motivated by a new study aiming to predict median property prices in 2018 at the small-area scale in Scotland, and details of this study are presented in Section 2. Section 3 provides a review of competitor prediction models, while our novel \texttt{CAR-Forest} algorithm is described in Section 4. The study design for assessing predictive performance is outlined in Section 5, while the study results are presented in Section 6. Finally, the paper ends in Section 7 with a summary of the main findings and areas for future work.

\section{Motivating study}
The aim of the study is to predict median property prices at the small-area scale in Scotland in 2018, which is the most recent year of data that are publicly available. The data relate to spatial units called Data Zones (DZ), which are a small-area geography containing between 500 and 1,000 people. Data Zones nest within 32 larger Local Authorities (LA), which are the administrative units that run public services such as schools and rubbish collections. Three of these LAs Na h-Eileanan Siar (Western Isles), Orkney, and Shetland are island communities that contain only 95 DZs in total, which are removed to avoid having small numbers of DZs in an LA when splitting the data into training and test sets. This leaves $N=6,881$ DZs as the study region, which comprise mainland Scotland and some of the islands. The data used in this study are described below, and unless otherwise stated were obtained from \url{https://statistics.gov.scot/home}.

\begin{figure}[p]
    \centering
	\begin{picture}(10,20)
	\put(-0.9, 10.5){\scalebox{0.13}{\includegraphics{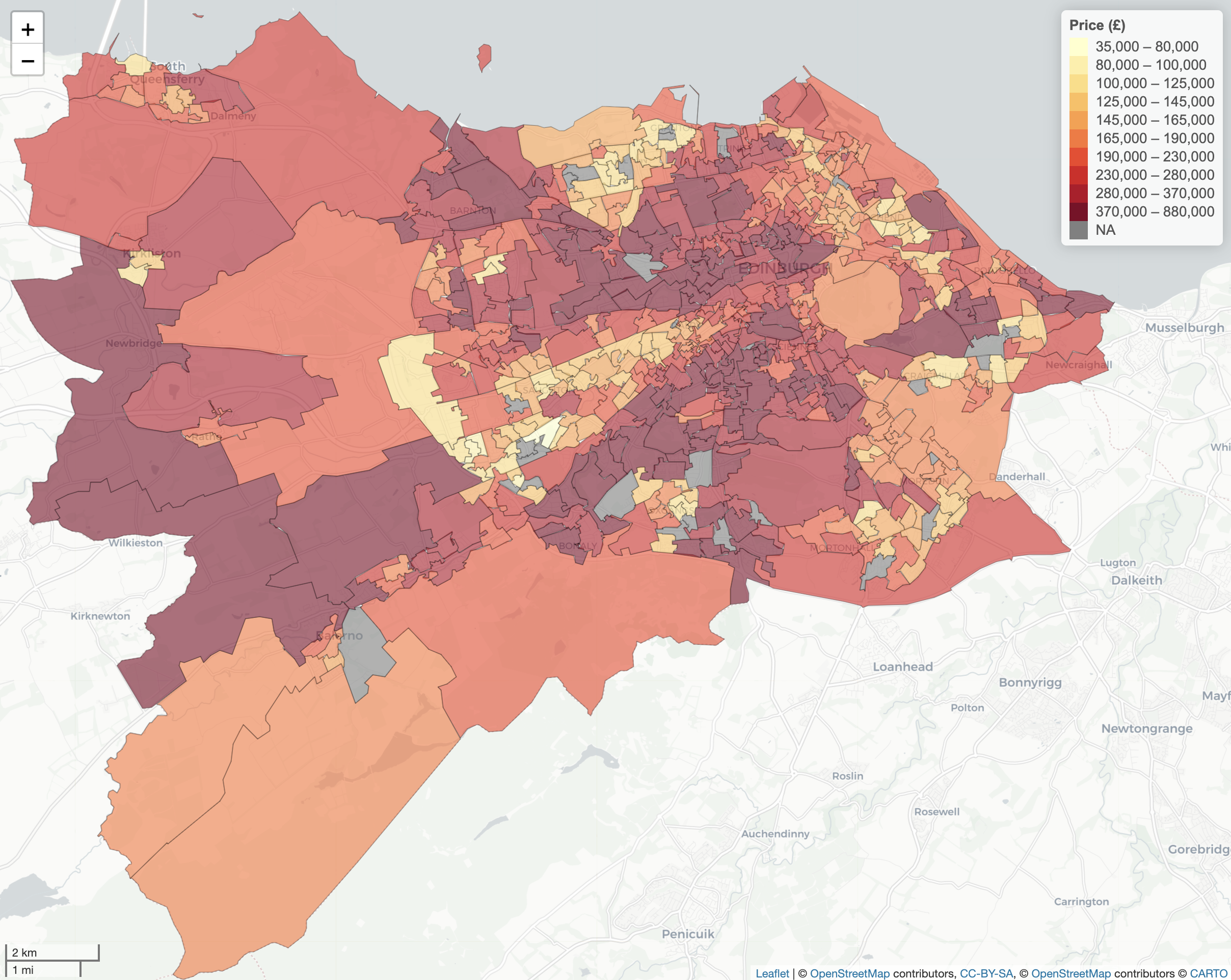}}}
	\put(-0.9, 0.2){\scalebox{0.165}{\includegraphics{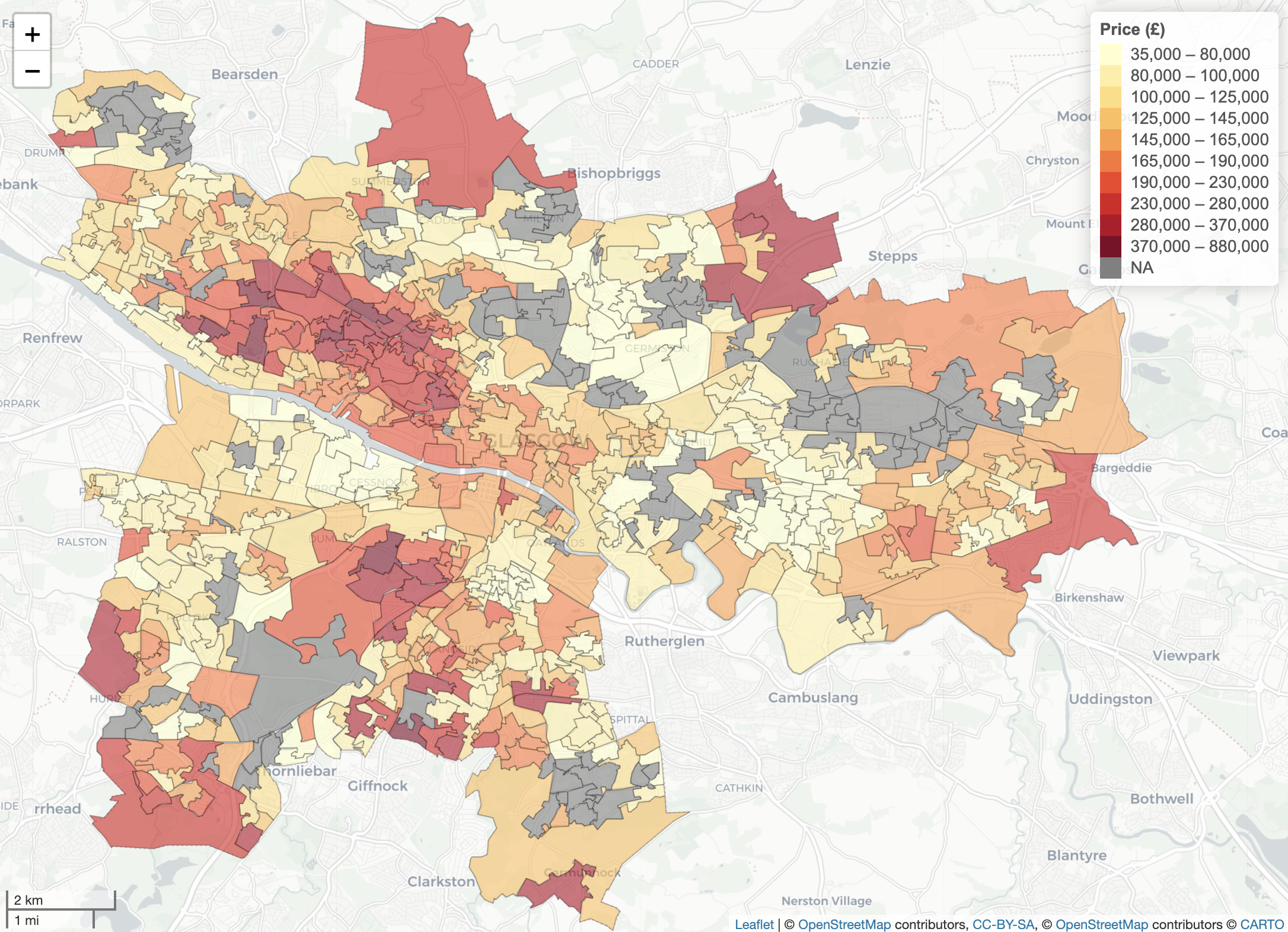}}}
    \put(-0.9, 19.5){\textbf{(A) -  Edinburgh}}
    \put(-0.9, 9.5){\textbf{(B) -  Glasgow}}
	\end{picture}
    \caption{Maps of median property prices in each DZ in Edinburgh (A, top) and Glasgow (B, bottom). Grey DZs have missing median property prices.}
    \label{figmaps1}
\end{figure}

\subsection{Target variable}
The target variable is the median selling price of all properties sold in 2018, with the median being used because it is robust to outlying observations. Median prices that are based on less than 5 sales are suppressed (or do not exist in the case of zero sales) to ensure individual properties are not identifiable, which results in around 9\% of DZs having missing values. Additionally, one DZ had a median price of just $\pounds$600, and as this is likely to be an error this value is treated as missing. The remaining data exhibit a skewed distribution that ranges between $\pounds 19,500$ and $\pounds 878,000$, with a median value of $\pounds 139,282$. Figure \ref{figmaps1} displays the spatial patterns in median property prices across the two largest cities of Edinburgh (A, top) and Glasgow (B, bottom), while all the data are not shown because most DZs would then be too small to see. The figure shows that  prices are more expensive in Edinburgh compared to Glasgow, with median prices of $\pounds 230,000$ and $\pounds 122,000$ respectively. Glasgow also exhibits a much higher proportion of DZs with missing property prices (shown in grey) than Edinburgh, being 16.8\% and 4.0\% respectively. These missing values appear to be spatially clustered in Glasgow but not in Edinburgh, with clusters in the residential areas of Drumchapel in the far north-west and Castlemilk in the far south. 

\subsection{Features}
A number of features that are likely to explain the spatial variation in median property prices were obtained, including characteristics of the DZ itself and the properties situated within them. Some of these features contain a small number of missing values, which are imputed using the K nearest neighbours (KNN) algorithm with $K=5$ as recommended by \cite{kuhn2019}. Additionally, a very small number of clear outliers were assumed to be data errors and imputed as above. The numeric features were then standardised to have a mean of zero and a standard deviation of one. The set of features is summarised below, with additional exploratory analysis given in Section 1 of the supplementary material.  

\subsubsection{Property characteristics}
Average property size is measured by the mean number of rooms excluding bathrooms and kitchens, while property type is summarised by the percentages of: (i) flats; and (ii) semi-detached / detached houses; in each DZ. Additionally, the density of properties is summarised by the number of dwellings per hectare. Finally, council tax is a levy paid by each householder for public services, and the council tax band of a property provides a crude measure of a property's worth. The latter has 8 levels labelled A to H, with the cheapest properties in band A and the most expensive in band H. The percentages of properties in each of these 8 bands is available, but as they are highly correlated principal components analysis (PCA) is applied to obtain independent features. The first 5 PCs explained over 95\% of the variation in these variables, and hence are used in the prediction model.

\subsubsection{Small-area characteristics}
The level of socio-economic deprivation in each Data Zone is measured by the Scottish Index of Multiple Deprivation (SIMD, \url{https://simd.scot/}), and we use data for the closest year available which is 2016. The SIMD is a composite index comprising 26 correlated indicators across the domains of access to services, crime, education, employment, health, housing and income, and a summary is presented in Section 1 of the supplementary material. The single indicator in the crime domain is removed because it has 435 (6\%) missing values, while the remaining  indicators have at most 14 missing values and are hence imputed. The income and employment domains each contain a single indicator which are used as-is, while the access to services domain is represented by the indicator summarising the drive time needed to reach a post office because the remaining indicators contain numerous outliers. The education, health and housing domains contain multiple correlated indicators, so PCA is again used to produce independent features. In each case enough PCs are kept to make the cumulative proportion of variation explained above 95\%, which resulted in the following number of features: education - 1,  health - 3,  housing - 2.

The urbanicity of each DZ is represented by an 8 fold urban-rural classification, which for simplicity is reduced to the following three levels: \texttt{urban} (Large Urban Areas, Other Urban Areas); \texttt{small-towns} (Accessible Small Towns, Remote Small Towns,  Very Remote Small Towns); and \texttt{rural} (Accessible Rural Areas, Remote Rural Areas, Very Remote Rural Areas). Here, \texttt{small towns} is treated as the baseline level, resulting in binary indicator variables for the \texttt{urban} and \texttt{rural} categories. The final features available comprise the local authority each DZ is contained within (a factor with 29 levels), and the easting (east-west) and northing (north-south) coordinates of the  centroid (central point) of each DZ. 

\subsection{Study aims}
Within the overarching aim of spatial areal unit prediction, this study addresses three key questions. Firstly, how does the predictive performance of the proposed \texttt{CAR-Forest} algorithm compare to a-spatial random forests, Bayesian CAR models and geographically weighted random forests? Secondly, how does property price predictability vary regionally across Scotland, and which areas can be predicted with the greatest and least amounts of accuracy? Thirdly,  what are the likely median property prices for the 9\% of  Data Zones that have missing values, and how do these predictions compare to the prices in the remaining Data Zones? This paper will thus provide users with information on average property prices in their local areas, as well as access to a state-of-the-art prediction algorithms for spatial areal unit data.

\section{Competitor prediction models}
This section briefly outlines the competitor prediction models that we benchmark our \texttt{CAR-Forest} algorithm against. In what follows, the study region is partitioned into $N$ non-overlapping areal units $\mathcal{S}=\{\mathcal{A}_1,\ldots,\mathcal{A}_N\}$, and median property prices are denoted by $\mathbf{Y}=[Y(\mathcal{A}_1),\ldots,Y(\mathcal{A}_N)]$. Additionally, $\mathbf{x}(\mathcal{A}_k)=[x_1(\mathcal{A}_k),\ldots,x_p(\mathcal{A}_k)]$ denotes a vector of $p$ features relating to $\mathcal{A}_k$. The $N$ areal units are randomly partitioned into a training set $\{\mathcal{A}_1,\ldots,\mathcal{A}_K\}$  and a test set  $\{\mathcal{A}_{K+1},\ldots,\mathcal{A}_N\}$, with each model being fitted to the training set and used for out-of-sample prediction on the test set.

\subsection{Normal linear model}
The simplest baseline model is the normal linear model, which when applied to the training set is given by  

\begin{eqnarray}
    Y(\mathcal{A}_k)&\sim&\mbox{N}\{\beta_0 +\mathbf{x}(\mathcal{A}_k)^{\top}\boldsymbol{\beta},\sigma^2\}\hspace{0.5cm}\mbox{for }k=1,\ldots,K,\label{linearmodel} 
\end{eqnarray}

where $(\beta_0, \boldsymbol{\beta}, \sigma^2)$ denote the intercept term, the $p\times 1$ vector of regression parameters and the variance parameter respectively. Parameter estimation is achieved using maximum likelihood, and further details are given by \cite{faraway2014} along with how to predict observations in the test set.

\subsection{Spatial conditional autoregressive model}
Residual spatial autocorrelation not accounted for by the covariates is ubiquitous in areal unit data, and  can be modelled by adding autocorrelated random effects to (\ref{linearmodel}). Conditional autoregressive (CAR, \citealp{besag1991}) priors are commonly specified for the random effects, with inference undertaken in a Bayesian setting, using either MCMC simulation (e.g., via \texttt{CARBayes}, \citealp{lee2013}) or integrated nested Laplace approximations (INLA, \citealp{rue2009}). Here we use INLA for its computational speed, due to the need to repeatedly fit the model multiple times when optimising tuning parameters.

CAR priors induce spatial dependence into the random effects via a (typically) binary neighbourhood matrix $\mathbf{W}$, whose $kj$th element $w_{kj}=1$ if areal units $\{\mathcal{A}_k, \mathcal{A}_j\}$ are spatially close and $w_{kj}=0$ otherwise ($w_{kk}=0~\forall k$). This specification implies that the random effects for areal units $\{\mathcal{A}_k, \mathcal{A}_j\}$ are partially autocorrelated if $w_{kj}=1$, but are conditionally independent given the remaining random effects if $w_{kj}=0$. Most commonly, $w_{kj}=1$ if $\{\mathcal{A}_k, \mathcal{A}_j\}$ share a common border and $w_{kj}=0$ otherwise. However, as we split the data into training and test subsets, some areal units will not share a border with any other units in these subsets, leading to an inappropriate specification of the CAR model. Therefore, for each subset we define $\mathbf{W}$ by the \emph{D nearest neighbours} algorithm, where $w_{kj}=1$ if $\mathcal{A}_j$ is one of the $D$ nearest neighbours to $\mathcal{A}_k$ in terms of inter-centroidal distance, and $w_{kj}=0$ otherwise. This creates an asymmetric neighbourhood matrix, which is made symmetric by setting $w_{jk}=1$ if initially ($w_{kj}=1$, $w_{jk}=0$). Here $D$ is a tuning parameter of the model that controls the structure of the spatial autocorrelation, with longer range partial autocorrelations being captured as $D$ increases. 

CAR priors have been proposed by \cite{besag1991}, \cite{leroux2000} and \cite{riebler2016}, and here we utilise the one proposed by \cite{leroux2000} because it only contains a single set of random effects $\{\phi(\mathcal{A}_1),\ldots, \phi(\mathcal{A}_K)\}$ whose level of spatial dependence is controlled globally by a single parameter $\rho$. Thus the full spatial CAR model applied to the training set is given by

\begin{eqnarray}
Y(\mathcal{A}_k) &\sim& \mbox{N}\{\beta_0+\mathbf{x}(\mathcal{A}_k)^{\top}\boldsymbol{\beta} + \phi(\mathcal{A}_k), \sigma^2\}\hspace{0.5cm}\mbox{for }k=1,\ldots,K\label{spatialcar}\\
\phi(\mathcal{A}_k)|\boldsymbol{\phi}(-\mathcal{A}_k),&\sim&\mbox{N}\left\{\frac{\rho\sum_{j=1}^{K}w_{kj}\phi(\mathcal{A}_j)}{\rho\sum_{j=1}^{K}w_{kj} + 1-\rho}, \frac{1}{\tau\left[\rho\sum_{j=1}^{K}w_{kj}+1-\rho\right]}\right\}\nonumber\\
\beta_0, \beta_j &\sim&\mbox{N}(0, 100000)\hspace{0.5cm}\mbox{for }j=1,\ldots,p\nonumber\\
\ln\left(\frac{\rho}{1-\rho}\right) &\sim& \mbox{N}(0, 100) \nonumber\\
\ln(\sigma^{-2}),~\ln(\tau) &\sim&  \mbox{log-gamma}(1, 0.01), \nonumber
\end{eqnarray}

where $\boldsymbol{\phi}(-\mathcal{A}_k)=\boldsymbol{\phi}\setminus\{\phi(\mathcal{A}_k)\}$. Here,  $\rho=0$ corresponds to spatial independence because then $\phi(\mathcal{A}_k)\sim\mbox{N}(0, 1/\tau)$, while if $\rho=1$ then (\ref{spatialcar}) becomes the intrinsic CAR prior for strong spatial autocorrelation proposed by \cite{besag1991}. Weakly informative prior distributions are specified for $(\beta_0, \boldsymbol{\beta}, \rho, \sigma^2, \tau)$ to let the data speak for themselves, which are the ones recommended by the INLA software used for inference \citep{rue2009}. Once fitted to the training set the model is used to predict property prices in the test set by sampling from the posterior predictive distribution 

\begin{equation}
f\{Y(\mathcal{A}_{K+1}),\ldots,Y(\mathcal{A}_N) |\tilde{\mathbf{Y}}\}=\int_{\boldsymbol{\Theta}}f\{Y(\mathcal{A}_{K+1}),\ldots,Y(\mathcal{A}_N)|\boldsymbol{\Theta}\}f\{\boldsymbol{\Theta}|\tilde{\mathbf{Y}}\}\text{d}\boldsymbol{\Theta},
\end{equation}

where $\boldsymbol{\Theta}$ denotes the set of model parameters and $\tilde{\mathbf{Y}}=[Y(\mathcal{A}_1),\ldots,Y(\mathcal{A}_K)]$ denotes the training data. Further details of the prediction are provided in Section 2.1 of the supplementary material accompanying this paper.

\subsection{Random forest model}
Random forests (RF) are one of the best performing machine learning prediction algorithms (\citealp{boehmke2020}), and were originally proposed by \cite{breiman2001}. They are based on the additive decomposition

\begin{eqnarray}
    Y(\mathcal{A}_{k})&=& m[\mathbf{x}(\mathcal{A}_k)] + \epsilon(\mathcal{A}_k)\hspace{1cm}\mbox{for }k=1,\ldots,N,
\end{eqnarray}

where $\{m[\mathbf{x}(\mathcal{A}_k)]\}$ are the true values and the errors $\{\epsilon(\mathcal{A}_k)\}_{k=1}^{N}$ across both training and test sets are assumed to be independent and identically distributed with some distribution $g(.)$. Random forests fit an ensemble of  $N_{tr}$ regression trees (\citealp{breiman1984}) to the training data $\{Y(\mathcal{A}_1),\ldots,Y(\mathcal{A}_K)\}$ to estimate $\{m[\mathbf{x}(\mathcal{A}_k)]\}$, and the predictions $\{\hat{m}[\mathbf{x}(\mathcal{A}_{K+1})],\ldots,\hat{m}[\mathbf{x}(\mathcal{A}_N)]\}$ of the test set  observations $\{Y(\mathcal{A}_{K+1}),\ldots,Y(\mathcal{A}_N)\}$ are the means of the predictions made by these $N_{tr}$ trees. Here we fix $N_{tr}=1,000$, which initial analyses showed was sufficient for the prediction error to stabilise. Each tree is fitted to an independent bootstrapped sample (with replacement) from the training data of the same size. A tree is built by a recursive binary partitioning algorithm, which considers a random subset of $m_{try}$ features when making each split in the tree. This recursive splitting continues until a stopping criterion is met, such as when making an additional split would result in a terminal node having less than $min_{node}$ observations. Full details of the algorithm are given in \cite{breiman1984}, while a practical introduction is given by \cite{boehmke2020}. Random forests only provide a single point prediction without a measure of predictive uncertainty, which we overcome using the  95\% prediction intervals for random forests proposed by \cite{zhang2020}. Further details are given in Section 2.2 of the supplementary material.

\subsection{Geographically weighted random forest model}
The most popular machine learning algorithm designed for spatially structured areal unit data is the geographical random forest (GRF, \citealp{georganos2021}), which is an extension of geographically weighted regression (\citealp{fotheringham2003}) to  a random forest context. It predicts average property price for a test set observation $Y(\mathcal{A}_r)$  by

\begin{equation}
  \hat{Y}(\mathcal{A}_r)  ~=~\alpha \hat{Y}^{local}(\mathcal{A}_r) + (1-\alpha)\hat{Y}^{global}(\mathcal{A}_r),
\end{equation}

where $\{\hat{Y}^{global}(\mathcal{A}_r), \hat{Y}^{local}(\mathcal{A}_r)\}$ respectively denote predictions from global (i.e., a standard random forest, see Section 3.3) and local random forest models, while $\alpha$ is a weight parameter in the interval $[0,1]$. The local random forest is constructed using only data from the $bw$ nearest areal units in the training set to $\mathcal{A}_r$, with the intuition being that these \emph{local} units are likely to be more similar to $\mathcal{A}_r$ and hence lead to an improved prediction of $Y(\mathcal{A}_r)$ compared to using areal units that are further away. This model thus has two additional tuning parameters compared to classical global random forests, namely $(bw, \alpha)$, which respectively control how localised the local random forest model is (bw) and the weight given to its prediction ($\alpha$). The original GRF algorithm proposed by \cite{georganos2021} did not provide measures of predictive uncertainty, which is rectified here using the approach proposed by \cite{zhang2020} that uses out-of-bag errors from the global random forest model. For details see Section 2.2 of the supplementary material.

\section{Methodology}
This section proposes a novel iterative spatial prediction algorithm for areal unit data called \texttt{CAR-Forest}, that uses random forests to estimate non-linear feature-target relationships and  Bayesian CAR models to allow for any residual spatial autocorrelation. Its rationale is outlined in Section 4.1, while algorithmic details are provided in Section 4.2.

\subsection{Overall approach and rationale}
The observed data $\{Y(\mathcal{A}_k)\}$ represent error-prone measurements of the true values $\{m[\mathbf{x}(\mathcal{A}_k)]\}$, leading to the decomposition

\begin{equation}
    Y(\mathcal{A}_k) ~=~ m[\mathbf{x}(\mathcal{A}_k)] + \epsilon(\mathcal{A}_k)~~~~\mbox{for }k=1,\ldots,N,
\end{equation}

for both the training and test sets. The errors $\{\epsilon(\mathcal{A}_k)\}$ are independent and identically distributed, and represent the noise in the observed data. The true values $\{m[\mathbf{x}(\mathcal{A}_k)]\}$ are unknown, and can be estimated using a random forest based on the available features $\{\mathbf{x}(\mathcal{A}_k)\}$.  Denoting this model-based estimate by $\{\hat{m}[\mathbf{x}(\mathcal{A}_k)]\}$,
the above decomposition becomes

\begin{equation}
    Y(\mathcal{A}_k) ~=~ \hat{m}[\mathbf{x}(\mathcal{A}_k)] + \{m[\mathbf{x}(\mathcal{A}_k)] - \hat{m}[\mathbf{x}(\mathcal{A}_k)] \}+ \epsilon(\mathcal{A}_k)~~~~\mbox{for }k=1,\ldots,N.
\end{equation}

The differences between the true values and the model estimates  $\{m[\mathbf{x}(\mathcal{A}_k)] - \hat{m}[\mathbf{x}(\mathcal{A}_k)]\}$ arise from an incorrect specification of the regression model, which is likely to be caused by a range of factors, including measurement error in the features and  unmeasured confounding. These unmeasured confounders are likely be spatially autocorrelated, and their exclusion from the model induces spatial autocorrelation into the differences $\{m[\mathbf{x}(\mathcal{A}_k)] - \hat{m}[\mathbf{x}(\mathcal{A}_k)]\}$. As these differences are unknown the standard approach is to replace them with a set of spatially autocorrelated random effects $\{\phi(\mathcal{A}_k)\}$, which yields the following general model:

\begin{equation}
    Y(\mathcal{A}_k) ~=~ \hat{m}[\mathbf{x}(\mathcal{A}_k)] + \phi(\mathcal{A}_k) + \epsilon(\mathcal{A}_k)~~~~\mbox{for }k=1,\ldots,N.
\end{equation}

This specification naturally suggests  a two-stage modelling cycle, where in the first stage the non-linear effects of the features on the target variable are estimated using a random forest. Then in the second stage the residual spatial autocorrelation not explained
by the features is modelled, by fitting a Bayesian CAR model to the target variable while treating the estimated feature effects $\{\hat{m}[\mathbf{x}(\mathcal{A}_k)]\}$ as fixed offsets. However, the presence of spatial-autocorrelation will likely affect the feature-target relationships estimated in stage 1. Therefore, we propose iterating these two stages  $r=1,\ldots, R$ times, where the random effects estimated in stage 2 are fed back into the random forest model in stage 1 as fixed offsets. The number of iterations $R$ is a tuning parameter of the algorithm, which together with the tuning parameters of the random forest ($m_{try}$, $min_{node}$) and the Bayesian CAR model ($D$)  are optimised via a cross-validation procedure (see Section 5). After $R$ iterations of these two steps the final Bayesian CAR model is used to make predictions for the test set. 

This approach leverages the best characteristics of both machine learning methods and spatial statistical models, namely the flexibility of the former for capturing non-linear relationships between a set of features and the target variable, and the ability of the latter for modelling  residual spatial autocorrelation. It thus extends random forest models for use in areal unit data applications where residual spatial autocorrelation is ubiquitous. We note that the GRF model proposed by \cite{georganos2021} does not allow for residual spatial autocorrelation directly, because it simply fits a-spatial random forest models to different local subsets of the training set.

\subsection{Implementation}
The iterative CAR-Forest prediction algorithm has the following tuning parameters: (i) the number of iterations of the algorithm $R$; (ii) the random forest specific tuning parameters $(m_{try}, min_{node})$; and (iii) the CAR model tuning parameter $D$. All of these are estimated using a 10-fold cross validation procedure applied to the training set, details of which are given in the next section. Thus the algorithm below is presented for a fixed set of tuning parameters. \vspace{0.2cm}

\hrule 
\texttt{Algorithm - CAR-Forest}
\hrule

\begin{description}
    \item\texttt{Stage 0: } Initialise the random effects by setting  $\phi(\mathcal{A}_k)=0$ for all training set observations, and fix the tuning parameters $(m_{try}, min_{node}, D, R)$.

    \item\texttt{Stage 1: } Iterate the following steps $r=1,\ldots, R$ times. 

    \begin{description}
        \item\textbf{A. } Compute the decorrelated target variable $Z(\mathcal{A}_k)=Y(\mathcal{A}_k) - \phi(\mathcal{A}_k)$ for observations in the training set $k=1,\ldots, K$.
        
        \item\textbf{B. } Fit a random forest model with tuning parameter $(m_{try}, min_{node})$ to the training set with features $\{\mathbf{x}(\mathcal{A}_k)\}_{k=1}^{K}$ and target variable $\{Z(\mathcal{A}_k)\}_{k=1}^{K}$ to estimate the effects of the features after adjusting for spatial autocorrelation. Use this model to produce out-of-bag predictions  $\{\hat{m}^{(k)}[\mathbf{x}(\mathcal{A}_k)]\}_{k=1}^{N}$ for both the training and test sets. 
        
        \item\textbf{C. } Fit the Gaussian Leroux CAR model described in Section~3.2
        \begin{equation}
        Y(\mathcal{A}_k) \sim \mbox{N}\{\beta_0 + \hat{m}^{(k)}[\mathbf{x}(\mathcal{A}_k)] + \phi(\mathcal{A}_k), \sigma^2\}\hspace{0.5cm}\mbox{for }k=1,\ldots,K,\nonumber
        \end{equation}

        to the training set, to produce an updated estimate of the residual spatial autocorrelation component $\{\phi(\mathcal{A}_k)\}_{k=1}^{K}$ via its posterior mean. The neighbourhood matrix $\mathbf{W}$ for the CAR model is constructed using the $D$ nearest neighbours rule. Here, the out-of-bag predictions $\{\hat{m}^{(k)}[\mathbf{x}(\mathcal{A}_k)]\}_{k=1}^{K}$ from Step B. are included in the model as a fixed offset. 
    \end{description}

    \item\texttt{Stage 2: } Use the final Gaussian CAR model from step C. obtained after $R$ iterations to produce predictions and 95\% prediction intervals for observations in the test set via their posterior predictive distributions. Note, these predictions also use the feature predictions $\{\hat{m}^{(k)}[\mathbf{x}(\mathcal{A}_k)]\}_{k=K+1}^{N}$ from the final random forest model. 
\end{description}
\hrule\vspace{1cm}

Further details about random forests (Stage 1 B.) and the Bayesian CAR model  used (Stage 1 C. and 2)  are provided in Section 3 of the main paper and Section 2 of the supplementary material. Steps B. and C. respectively produce and use out-of-bag predictions $\{\hat{m}^{(k)}[\mathbf{x}(\mathcal{A}_k)]\}_{k=1}^{K}$ from the random forest for the training set, which for areal unit $\mathcal{A}_k$ are made by averaging the predictions from the sub-forest of trees that were fitted without using $Y(\mathcal{A}_k)$. Out-of-bag predictions are used so that the training and test set predictions are generated in the same way, i.e., without using the data point in question. If one instead replaced $\{\hat{m}^{(k)}[\mathbf{x}(\mathcal{A}_k)]\}_{k=1}^{K}$ with in-sample fitted values, then they would likely be closer to the observed data compared to those in the test set, leading to an underestimation in predictive uncertainty. The above algorithm is implemented in \texttt{R}, and software allowing others to apply the method to their own data is available in Section 3 of the supplementary material. The software uses the \texttt{ranger} (\citealp{wright2017}) package to fit the random forests, and the \texttt{INLA} package (\citealp{rue2009}) to fit the Gaussian CAR models.

\section{Study design for assessing predictive performance}
The study design for assessing the predictive performance of the CAR-Forest algorithm and its competitors for the property price motivating study is outlined below, while a simulation study evidencing its performance in a range of controlled settings is presented in Section 4 of the supplementary material. The predictive performances of the following five models are compared: (i) a normal linear model (\texttt{LM} - Section 3.1); (ii) a spatial CAR model (\texttt{CAR} - Section 3.2); (iii)  a random forest model (\texttt{RF} - Section 3.3); (iv) a geographically weighted random forest model (\texttt{GRF} - Section 3.4); and (v) the CAR-Forest algorithm proposed in Section 4 (\texttt{CAR-Forest}). Note, we also compared a simplified non-iterative form of the CAR-Forest algorithm that is equivalent to $R=1$, but as it did not perform as well as the full CAR-Forest algorithm its results are not shown for brevity. The normal linear model is included for its simplicity, while the remaining 3 competitors to CAR-Forest comprise state of the art models in spatial statistics, machine learning and an existing fusion of these paradigms. Predictive performance is assessed by splitting the  $N=6,264$ Data Zones that contain non-missing median property price values into an 80\% training set (5,011) and a 20\% test set (1,253), which is repeated 5 times to ensure the results are not affected by the particular choice of training-test split. 

With the exception of the normal linear model the remaining models contain tuning parameter(s), which 
are initially optimised using the training set. This optimisation is done using a 10-fold cross validation procedure, which splits the training set into ten random subsets of approximately equal size. Each model is fitted to nine of these subsets with different combinations of tuning parameters, and for each combination the observations in the tenth subset, known as the validation set, are predicted. This process is repeated treating each of the ten subsets as the validation set once, and the optimal values of the tuning parameters are the combination that minimise the root mean square error (see below for the definition) of the predictions. This process is repeated independently for each of the five training and test splits. 

Once the optimal tuning parameters have been chosen, each model is refitted to the full training set using these optimal values, which are then used to make out-of-sample predictions for the test set. As median property price is a continuous measurement, the quality of these predictions is assessed using the following standard metrics. In what follows $\{Y(\mathcal{A}_r), \tilde{Y}(\mathcal{A}_r)\}$ denote the observed median property price and the prediction for the $r$th areal unit in the test set respectively, where following the notation in Section 3 $r=K+1,\ldots,N$. 

\begin{eqnarray}
    \text{\textbf{Root mean square error}} &-& \text{\textbf{RMSE}}=\sqrt{\frac{1}{N-K}\sum_{r=K+1}^{N}\left[\tilde{Y}(\mathcal{A}_r)-Y(\mathcal{A}_r)\right]^2}.\nonumber\\
    \text{\textbf{Median absolute error}} &-& \text{\textbf{MAE}}=\mbox{Median}_{r=K+1,\ldots,N}\left\{\left|\tilde{Y}(\mathcal{A}_r)- Y(\mathcal{A}_r)\right|\right\}.\nonumber\\
\text{\textbf{Coverage probability}} &-& \text{\textbf{CP} = The proportion of the $N-K$ 95\% prediction} \nonumber\\
&&\text{intervals that contain the true value.}\nonumber\\
\text{\textbf{Average interval width}} &-& \text{\textbf{AIW} = The average width of the $N-K$ 95\%} \nonumber\\
&&\text{prediction intervals.}\nonumber
\end{eqnarray}

The accuracy of the point prediction is summarised by both the RMSE and MAE metrics, with the best model minimising both quantities. We present both metrics because as the RMSE utilises both arithmetic mean and squared operators it is much less robust to individual DZs with big prediction errors than the MAE is. The appropriateness of the 95\% prediction intervals is quantified by the coverage probability and average interval width, and the former should be close to 0.95 if predictive uncertainty is appropriately captured. Finally, the average interval width should be as small as possible as long as the coverage probability is close to 0.95.

\section{Results from the property price study}
This section presents the results of the motivating study, focusing on the 3 questions outlined in Section 2.3. Section 6.1 describes the implementation of the models, while Section 6.2 compares their predictive abilities. Section 6.3 provides a local authority comparison of predictive performance, while Section 6.4 predicts median property prices for the Data Zones with missing values.  

\subsection{Model implementation}
All models are applied with the complete set of features described in Section 2, which initial analyses showed performed similarly to using an additional forwards or backwards stepwise feature selection approach. The easting and northing coordinates of each Data Zone's centroid are included as features in the linear model, random forest and geographical random forest because these models have no other way of capturing spatial location, while this is not necessary for the CAR and the CAR-Forest approaches that utilise spatial random effects. 

Initially, the normal linear model was fitted to both median property price and its natural logarithm, and as the residual normality assumption is only plausible on the log scale, all models are applied on this scale in the interests of fairness. The resulting predictions are then back-transformed to the original scale when computing the predictive performance metrics outlined in the previous section. As a number of the models are based on a normality assumption, the back transformation for all point predictions follows the log-normal result that if $\mathbf{X}\sim\mbox{N}(\mu, \sigma^2)$, then $\mathbf{E}[\exp(X)]=\exp(\mu + \sigma^2/2)$. Here, the linear, CAR and CAR-Forest algorithms provide estimates of $\sigma^2$ directly from the model, while for the random forest and its geographical extension $\sigma^2$ is estimated by the sample variance of the out-of-bag prediction errors following the ideas in \cite{zhang2020}. Note, a simple exponential back-transformation resulted in biased predictions in all cases, where as the above transformation leads to negligible bias. 

The CAR model has a single tuning parameter $D$ that determines the construction of the neighbourhood matrix $\mathbf{W}$, and the values considered comprise $D\in\{3,5,7,9\}$. All models that incorporate a random forest component are implemented with $N_{tr}=1,000$ trees, which initial analyses showed was sufficient for the prediction error to stabilise.  The random forest model was optimised with respect to all possible combinations of the tuning parameters $m_{try}\in\{10, 20, 30, 40, 51\}$ and $min_{node}\in\{1,5,10\}$, where the largest value of $m_{try}$ is chosen to be equivalent to bagging. The GRF model was optimised with respect to the same set of $m_{try}$ and $min_{node}$ values, as well as the additional parameters $(bw, \alpha)$. Following \cite{georganos2021} these latter parameters were optimised over all possible combinations of $bw\in\{100, 500, 1,000\}$ and $\alpha\in\{0.25, 0.5, 0.75, 1\}$. Finally, the CAR-Forest algorithm was optimised with respect to the tuning parameters from the CAR model ($D$), the random forest model ($m_{try}$, $min_{node}$) and the total number of iterations of the algorithm ($R$). The same sets of possible values described above for the first two type were considered, while we considered all possible combinations of $R=1,\ldots,5$. 

\begin{table}[b]
    \centering
    \begin{tabular}{lrrrrrr}\hline
   \multirow{2}{*}{\textbf{Model}} & \multicolumn{6}{c}{\textbf{Split}} \\
       &\textbf{1}   & \textbf{2}   &\textbf{3}   &\textbf{4}     &\textbf{5} &\textbf{Mean}  \\\hline
   \textbf{RMSE}   &&&&&&\\
\texttt{LM} & $\pounds$46,992 & $\pounds$50,132 & $\pounds$49,569 & $\pounds$45,879 & $\pounds$42,331 & $\pounds$46,981\\
\texttt{CAR} & $\pounds$41,617 & $\pounds$44,528 & $\pounds$45,084 & $\pounds$40,845 & $\pounds$37,115 & $\pounds$41,838\\
\texttt{RF} & $\pounds$41,222 & $\pounds$46,407 & $\pounds$44,988 & $\pounds$42,622 & $\pounds$36,884 & $\pounds$42,424\\
\texttt{GRF} &$\pounds$40,562 & $\pounds$45,713 & $\pounds$44,640 & $\pounds$41,218 & $\pounds$35,288 & $\pounds$41,484\\ 
\texttt{CAR-Forest} & $\pounds$40,134 & $\pounds$43,585 & $\pounds$42,400 & $\pounds$39,798 & $\pounds$35,600 & $\pounds$40,303\\  \hline

   \textbf{MAE}   &&&&&&\\
\texttt{LM} & $\pounds$20,401 & $\pounds$19,816 & $\pounds$20,718 & $\pounds$19,753 & $\pounds$18,988 & $\pounds$19,935\\
\texttt{CAR} & $\pounds$18,565 & $\pounds$17,519 & $\pounds$17,614 & $\pounds$17,577 & $\pounds$17,388 & $\pounds$17,733\\
\texttt{RF} & $\pounds$18,124 & $\pounds$17,736 & $\pounds$18,255 & $\pounds$18,588 & $\pounds$17,181 & $\pounds$17,977\\
\texttt{GRF} & $\pounds$17,680 & $\pounds$17,352 & $\pounds$17,600 & $\pounds$18,585 & $\pounds$16,513 & $\pounds$17,546\\ 
\texttt{CAR-Forest} & $\pounds$16,684 & $\pounds$16,546 & $\pounds$16,997 & $\pounds$16,546 & $\pounds$16,344 & $\pounds$16,623\\ \hline

   \textbf{CP}   &&&&&&\\
\texttt{LM} & 0.932 & 0.943 & 0.936 & 0.943 & 0.950 & 0.941\\
\texttt{CAR} & 0.947 & 0.947 & 0.938 & 0.946 & 0.949 & 0.945\\
\texttt{RF} & 0.935 & 0.951 & 0.952 & 0.944 & 0.960 & 0.949\\
\texttt{GRF} & 0.947 & 0.954 & 0.954 & 0.951 & 0.966 & 0.955\\ 
\texttt{CAR-Forest} &0.935  &0.947  &0.939  &0.943  &0.958  &0.944 \\\hline

   \textbf{AIW}   &&&&&&\\
\texttt{LM} & $\pounds$155,091 & $\pounds$148,898 & $\pounds$152,074 & $\pounds$151,954 & $\pounds$153,574 & $\pounds$152,318\\
\texttt{CAR} & $\pounds$148,860 & $\pounds$147,725 & $\pounds$150,017 & $\pounds$146,681 & $\pounds$147,125 & $\pounds$148,082\\
\texttt{RF} & $\pounds$148,397 & $\pounds$150,477 & $\pounds$151,874 & $\pounds$148,213 & $\pounds$150,840 & $\pounds$149,960\\
\texttt{GRF} & $\pounds$150,160 & $\pounds$148,381 & $\pounds$152,490 & $\pounds$149,820 & $\pounds$151,432 & $\pounds$150,457\\ 
\texttt{CAR-Forest} & $\pounds$142,975 & $\pounds$139,801 & $\pounds$139,350 & $\pounds$141,647 & $\pounds$140,602 & $\pounds$140,875\\ \hline
    \end{tabular}
    \caption{Comparison of the out-of-sample predictive abilities of each model for each data split, and the overall mean over all 5 splits. The acronyms in the table denote: RMSE - root mean square error; MAE - median absolute error; CP - coverage probability; and AIW - average interval width.}
    \label{table_predictions}
\end{table}

\subsection{Comparing the predictive ability of the models}
The predictive abilities of the models for each training and test split as well as the mean over all five splits are summarised in Table \ref{table_predictions}, which presents the four metrics outlined in Section 5.2, namely RMSE, MAE, CP and AIW. The CAR-Forest algorithm has the best average point prediction in terms of both RMSE and MAE over all five training-test splits, with mean reductions compared to the best existing model of $\pounds$1,181 (RMSE) and $\pounds$923 (MAE) respectively. The GRF model performs the best out of the existing models in both RMSE and MAE, while the spatial CAR model outperforms an a-spatial random forest. The 95\% prediction intervals from all six models exhibit close to their nominal coverage probabilities (CP), with values ranging between 0.941 (linear model) and 0.955 (GRF) on average across all 5 training-test splits. The CAR-Forest intervals are $\pounds$7,207 narrower on average compared to those from the best existing model (CAR), suggesting that it provides the most precise inference from the set of models compared in addition to its superior point prediction.

\subsection{Comparing the accuracy of property price predictions by local authority}
\begin{figure}[t]
    \centering
\scalebox{0.2}{\includegraphics{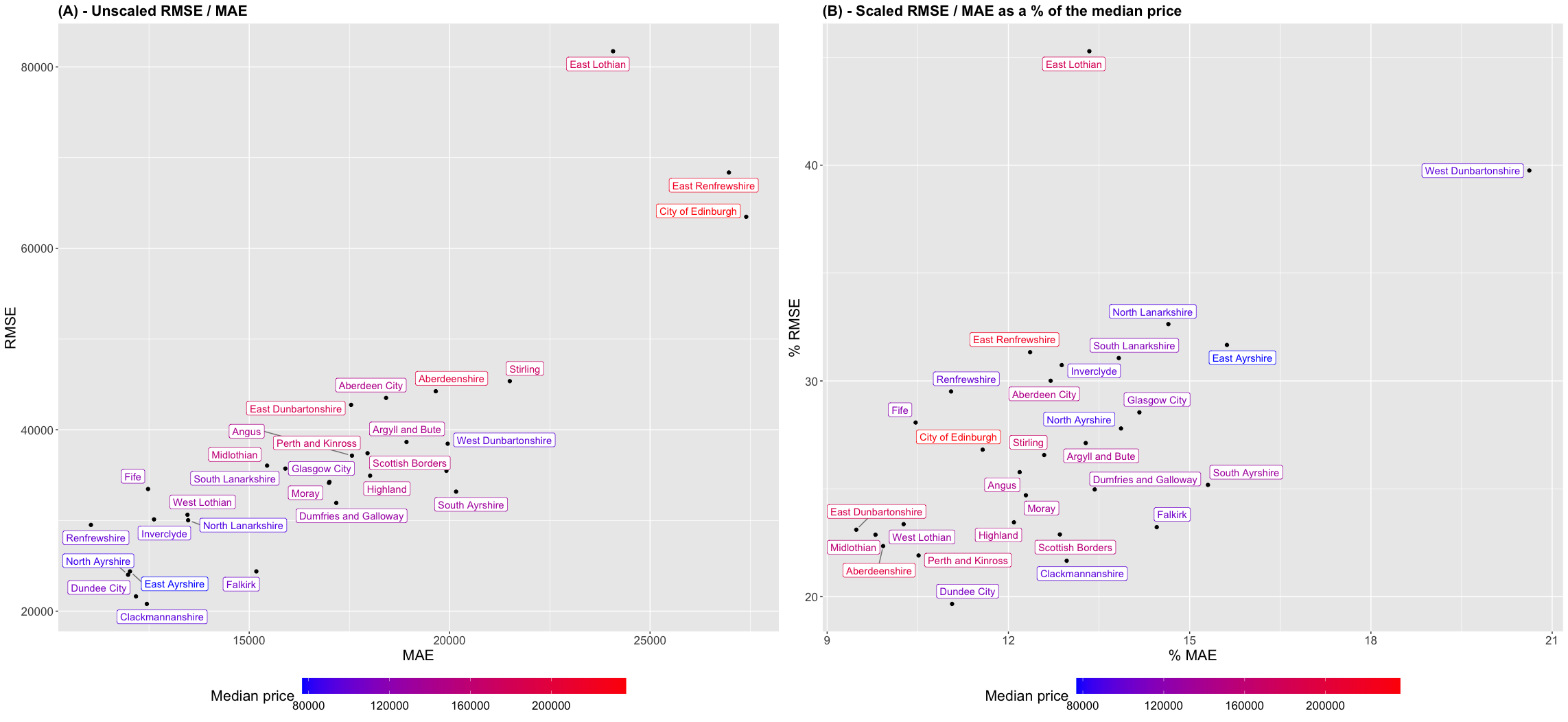}}
    \caption{Comparison of the point prediction accuracy of the CAR-Forest algorithm by local authority, as measured by MAE and RMSE. Panel (A) presents the un-scaled MAE / RMSE values, while panel (B) presents metrics scaled by the median property price in each LA.}
    \label{figLA}
\end{figure}

To get a regional view of the predictability of property prices in Scotland, we compute the RMSE and MAE metrics for the CAR-Forest model predictions separately for each of the 29 local authorities in the study. These regional metrics are plotted against each other in Figure \ref{figLA}, which presents the mean values over all 5 training and test splits. The colours of the labels in the figure denote the median price across DZs in each LA, with blue and red respectively representing cheaper and more expensive regions to live. The left panel (A) in Figure \ref{figLA} displays the absolute RMSE and MAE values, while the right panel (B) presents both metrics as  percentages of their median  property prices to remove the scaling effect caused by some regions having more expensive properties than others. 

Both panels show fairly strong linear relationships between the RMSE and the MAE metrics, suggesting that the relative accuracy of property price predictions by local authority are similar regardless of which metric is used. The left panel (A) shows that LAs that have more expensive properties on average  have higher prediction errors, with the City of Edinburgh, East Lothian and East Renfrewshire having the least accurate predictions in absolute terms, while Clackmannanshire and Dundee City have the most accurate predictions. However, once these prediction metrics have been scaled to account for differences in median property prices (panel (B)), then East Lothian and West Dumbartonshire  have the least predictable prices, while Aberedeenshire, Dundee City and East Dumbartonshire are the most predictable. These results also show there are no clear spatial or urban-rural trends in the relative predictability of Scotland's housing market, as nearby or similar areas do not necessarily have similar prediction metrics.

\subsection{Predicting missing median property prices}
As CAR-Forest is the best performing prediction algorithm in this study, it is re-fitted to 
all Data Zones that have property price data and used to predict the remaining DZs with missing values. The model is fitted with $N_{tr}=1,000$ trees as before, and we use $R=5$ because this value was optimal in all five training-test splits in the tuning parameter optimisation phase. The remaining tuning parameters are chosen as the mode of the optimal values across the five training-test splits, which results in $D=7$, $m_{try}=51$  and $min_{node}=5$. 

\begin{figure}
    \centering
\scalebox{0.4}{\includegraphics{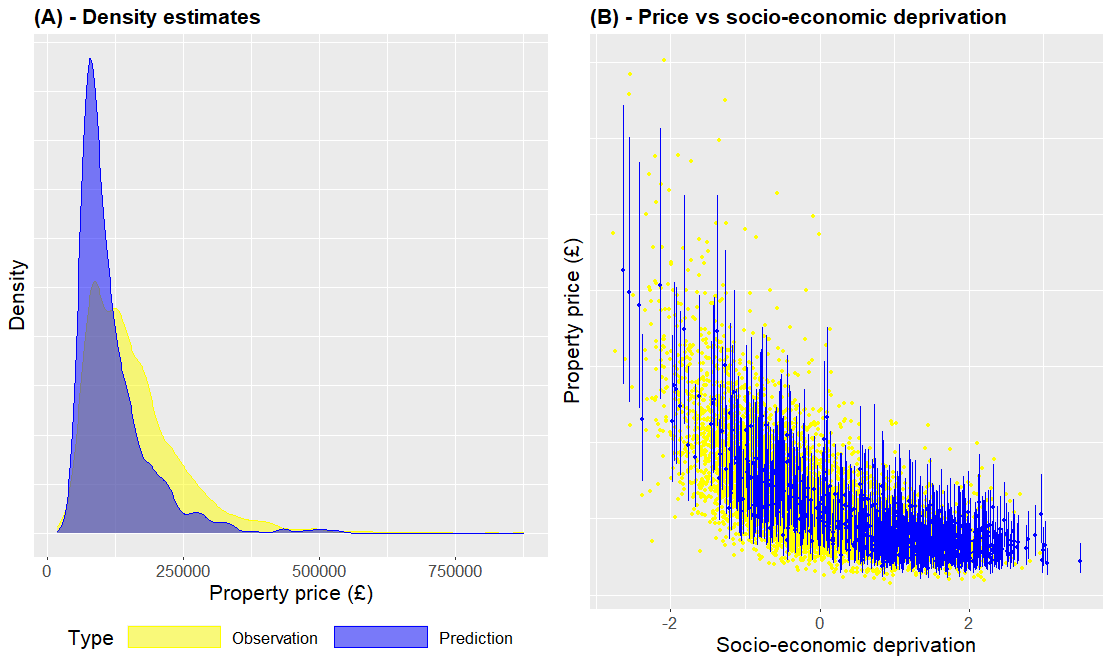}}
    \caption{Comparison of the Data Zones with predicted (blue - predictions and 95\% prediction intervals) and observed (yellow - observations) property prices. The left panel (A) shows density estimates while the right panel (B) shows the relationship between price / prediction and socio-economic deprivation as measured by the first principal component of the education domain of the SIMD.}
    \label{figpred}
\end{figure}

The distributions of both the predictions (617 DZs) and the observed price data (6,264 DZs) are displayed in panel (A) of Figure \ref{figpred} by density estimates, with the predictions in blue while the observed data are in yellow. The figure shows that the two distributions are skewed to the right with broadly similar shapes, but that the Data Zones with missing values have lower average prices (median $\pounds$96,335) compared to those with available data (median $\pounds$139,282). The likely reason for this price differential is illustrated in panel (B) of Figure  \ref{figpred}, which presents the predictions (blue, point estimates and 95\% prediction intervals) and observed prices (yellow, observed values) against a measure of socio-economic deprivation, specifically the first principal component of the education domain of the SIMD. The figure shows that property prices decrease as the level of socio-economic deprivation increases as expected, and that a high proportion of the Data Zones with missing price data are socio-economically deprived.

\section{Discussion}
This paper has proposed the first fusion of random forests and conditional autoregressive models for the prediction of spatial areal unit data with missing values, and has evidenced its improved predictive performance compared with a number of state-of-the-art alternatives. This improvement relates to both the accuracy of its point predictions as measured by RMSE and MAE, and the improved precision of its predictions as measured by their narrower 95\% prediction intervals that retain close to the nominal coverage levels. This superior performance is likely to be because it combines the best aspects of both random forests and CAR-based models, namely the flexibility of the former's  feature-target relationships and the ability of the latter to model residual spatial autocorrelation via random effects. This paper also provides the first coherent framework for applying random forests to spatially autocorrelated areal unit data, because unlike the RFRK (\citealp{hengl2015}) and GLS (\citealp{saha2023}) algorithms for point-level data, existing areal unit level methods such as geographical random forests (\citealp{georganos2021}) do not explicitly allow for the residual spatial autocorrelation in the data after the feature effects have been accounted for. 

In the motivating property price study the magnitude of the improved predictive performance of the CAR-Forest algorithm compared to the best competitor is around 2.8\% (around $\pounds$1,181) in RMSE, 5.3\% (around $\pounds$923) in MAE, and 4.9\% (around $\pounds$7,207) in the precision of its 95\% prediction intervals. These improvements are mainly benchmarked against the geographical random forest model, which generally outperforms the linear, random forest and CAR models in our study. Finally, a comparison of the random forest and CAR models shows that the latter slightly outperforms the former in terms of predictive performance, suggesting that for these property price data the ability to capture residual spatial autocorrelation is more important than capturing non-linear feature-target relationships. 

The simulation study presented in the supplementary material illustrates that the models perform as one would expect under controlled conditions, which provides further evidence about the utility and robustness of the CAR-Forest algorithm. Firstly, in the presence of both non-linear feature-target relationships and residual spatial autocorrelation CAR-Forest outperforms all the competitors, because it is the only method that can accommodate both of these components. Secondly, CAR-Forest does not perform as well as the CAR model when all the feature-target relationships are exactly linear, which is because in this case the additional unnecessary flexibility of the random forest results in poorer predictive performance compared with rigidly enforcing the relationships to be linear. Thus, in practical applications one could identify any features in advance that exhibit exactly linear relationships with the target variable via exploratory scatterplots, and include those features as linear terms in the CAR component of the model rather than in the random forest.

This paper has focused exclusively on predictive performance as the single inferential goal. It therefore opens up a new field of research at the intersection of machine learning and spatial statistics, which includes both methodological and application orientated challenges. From a methodological perspective we have only considered a fairly simple spatial data structure, which leads to the natural question of how one should extend the methods developed here to a multivariate spatio-temporal data context. In this setting different random forest models would be needed for each target variable and possibly time period, while complex residual autocorrelations over space, time and between the target variables would need to be accounted for. 

A second challenge concerns the inferential goal, which here was restricted to the prediction of areal units with missing data values. In this context the effects of the features on the target variable were not of direct interest, which allowed us to plug their estimated effects from the random forest into the second stage CAR model as a fixed offset. However, in the field of spatial ecological regression the relationship between a feature(s) and the target variable  is the primary inferential goal, meaning that the effects of the features and their uncertainty needs to be quantified.  One possible solution is to split the features into a set of those that are of primary interest and a second set of confounders, with the confounders remaining in the random forest component due to its flexibility while the features of primary interest are included in the Bayesian CAR model as suggested above in the context of linear feature-target relationships. A second possible solution is to use interpretable machine learning tools for feature effects such as partial dependence plots (\citealp{friedman2001b}) and variable importance plots (see \citealp{greenwell2020}).

Finally,  as noted in the introduction spatial areal unit data are prevalent in many fields, which gives the  CAR-Forest algorithm a wide range of possible application areas. One important such area is disease mapping (\citealp{lawson2018}), where the fusion of machine learning and spatial autocorrelation models has the potential to help answer a number of public health questions, such as where are the hotspots of disease risk,  which features affect disease risk, and how big are the health inequalities and how are they changing over time.

\section*{Acknowledgements}
For the purpose of open access, the author has applied a Creative Commons Attribution (CC BY) licence to any Author Accepted Manuscript version arising from this submission.

\bibliographystyle{imsart-nameyear} 
\bibliography{MDL_AOASsubmission}       


\end{document}